\documentclass[prd,a4paper,twocolumn]{revtex4}

\usepackage{graphicx}
\usepackage{psfrag}
\usepackage{inputenc}
\inputencoding{latin9}

\newcommand{\nc}{\newcommand}

\def\lsim{\; \raise0.3ex\hbox{$<$\kern-0.75em
      \raise-1.1ex\hbox{$\sim$}}\; }
\def\gsim{\; \raise0.3ex\hbox{$>$\kern-0.75em
      \raise-1.1ex\hbox{$\sim$}}\; }

\nc{\be}[1]{\begin{equation}\mbox{$\label{#1}$}}
\nc{\bea}[1]{\begin{eqnarray} \mbox{$\label{#1}$}}
\nc{\Section}[2]{\section{#2}\label{#1}}
\nc{\Bibitem}[1]{\bibitem{#1}}
\nc{\Label}[1]{\label{#1}}

\nc{\eea}{\end{eqnarray}}
\nc{\ee}{\end{equation}}

\begin{document}

\title{Complex supergravity quintessence models 
confronted with Sn Ia data}

\author{K. Henttunen}
\thanks{kajohe@utu.fi}
\affiliation{Department of Physics,
University of Turku, FIN-20014 Turku, FINLAND}
\author{T. Multam\"aki}
\thanks{tuomul@utu.fi}
\affiliation{Department of Physics,
University of Turku, FIN-20014 Turku, FINLAND}
\author{I. Vilja}
\thanks{vilja@utu.fi}
\affiliation{Department of Physics,
University of Turku, FIN-20014 Turku, FINLAND}
\date{\today}

\begin{abstract}
A class of supergravity inspired quintessence models
is studied by comparing to cosmological data. The set of considered
models includes several previously studied quintessential 
potentials, as well as the $\Lambda$CDM model.
We find that even though the commonly studied supergravity 
inspired quintessence models fit the data better than the
$\Lambda$CDM model, they are a relatively 
poor fit when compared to the best fit model in the studied class.
Our results suggest a low energy scale, less than 
${\cal{M}}\sim 1$ TeV, for the effective supergravity potential.

\end{abstract}

\maketitle

\section{Introduction}

According to the cosmic microwave background radiation (CMB), supernova and 
large scale structure experiments 
\cite{Spergel:2003cb,Riess:1998cb,Perlmutter:1998np,Peacock:2001gs,
Tegmark:2003ud}, 
the universe is nearly flat and accelerating on the scales of the present 
cosmological horizon.
Therefore, it appears, either that the total energy density of the universe is
currently dominated by an dark energy component, or 
that the Einsteinian gravity needs to be modified at large distances.
Within Einsteinian gravity, the constant solution for dark energy is the vacuum
energy model ($\Lambda$CDM) (with constant $\rho_{\Lambda}$ and $\omega_{\Lambda}=\frac{p}{\rho}=-1$, 
see \cite{Krauss:1995yb} and references within).
The $\Lambda$CDM model is widely considered as a cosmological
concordance model
as it generally explains the present cosmological observations.
The observed value for the vacuum energy $\rho_\Lambda\sim\rho_C\sim 10^{-47}$ GeV$^4$ is, however, unnaturally small for constant vacuum energy model.
The underlying models of particle physics can not provide a natural 
explanation to the necessity of careful fine-tuning of the energy scale.
Neither does the vacuum energy model explain why the dark energy domination started just recently; {\it i.e.} why the energy densities of matter and dark energy coincide today, although these two energy densities have evolved differently throughout the history of the universe.
Had the dark energy domination started earlier, present structures could not have been formed; if later, no acceleration could be presently observed.
It seems, as the initial conditions have to be very carefully fine-tuned to produce this coincidence. Those who are not content with the anthropic principle (see {\it e.g.} \cite{Weinberg:1988cp}), need to find a quantitative explanation to this problem.

The quintessential scenario represents a class of dark energy models that are
able to produce negative pressure. From this point of view, the accelerated 
expansion is explained with a minimally coupled dynamical scalar field that is 
evolving along a suitable potential. 
A tracking \cite{Steinhardt:1999nw} feature of a quintessence model, allows the quintessence field to mimic the evolution of the background fluid until very recent times, when it becomes the dominant component in the total energy density. Thus the quintessential model with a tracking property opens up a possibility to explain why a wide range of initial field values and energy scales converge to a recent epoch of dark energy domination.
Furthermore, the tracking feature is a favourable feature when considering inflationary models. A wide range of post inflationary conditions naturally converge to a suitable late time cosmology.
This property makes the tracking quintessence an appealing alternative to the cosmological constant model.

Various kinds of quintessence potentials exist. A class based on high energy physical considerations is studied in this paper.
We consider a general complex quintessential model based on an effective
supergravity model by fitting to current SN Ia data.

\section{Model}

The dark energy model discussed here is a complex quintessence model based on an effective 
supergravity model on a Friedmannian background.
The supergravity model introduced in \cite{Brax:1999gp} by Brax and Martin 
(hereafer BM), naturally leads to quintessence potential $\propto  
e^{-|\psi |^\beta}/ |\psi |^\alpha$ with the simplest K\"ahler potential. However, with a suitable choice of the effective K\"ahler potential, the BM 
model can be extended to a class of potentials covering a number of 
well studied quintessence potentials:
\be{pot}
V(\psi) = {{\cal M}^{\alpha+4} \over |\psi |^\alpha }e^{\left (\frac \kappa 2 |\psi |^2\right )^{\beta/2}},
\ee
where $\kappa=8\pi G_N$, $\beta$ and $\alpha$ are positive integers and ${\cal M}$ is the energy scale of the potential.
This potential includes several well studied dark energy models
as special cases: the $\Lambda$CDM model ($\alpha=0,\ \beta=0$),
the BM model ($\beta=2, \alpha=11$), inverse power \cite{Ratra:1987rm} ($\beta=0$) and pure exponential potential \cite{Rubano:2001su,Barreiro:1999zs} ($\alpha=0$) models. 

These special cases have usually been studied with a real scalar field.
A real scalar field is not a natural part of the supergravity model, but merely a special choice. Also for other types of quintessence potentials 
there is no {\it a priori} reason to restrict the study to a real field either.
Therefore, in this paper we consider a versatile potential with a 
complex field $\psi$.

The equation of motion of the complex scalar field 
$\psi=\phi e^{i\theta}$ moving in this potential in a cosmological setting
is 
\be {dyna}
0=
\ddot{\phi}+2i\dot{\theta}\dot{\phi}+i\ddot{\theta}\phi-\dot{\theta}^2\phi
+3H(\dot{\phi}+i\dot{\theta}\phi)+\frac{dV_\phi(\phi^2)}{d\phi^2}\phi,
\end{equation}
where $H$ is the Hubble parameter.
Considering the imaginary part of Eq. (\ref{dyna}) it is evident that we
can define a constant of motion, a conserved charge $L=\dot{\theta}\phi^2a^3$.
Using this to replace $\dot{\theta}$ in the real part of Eq. (\ref{dyna}) results in an equation of motion for $\phi$ only.

\section{Cosmology}

It is widely accepted, that the energy content of the universe can be 
well described with a model of several interacting perfect fluids 
(\emph{i.e.} radiation, neutrinos, baryons,  cold dark matter and dark energy).
The dynamics of the universe is here described by the conventional Friedmannian cosmology:
\begin{equation}
H^2+\frac k{a^2}=\frac{8\pi G}{3}\Big(\rho_M+\rho_R+\rho_Q\Big),
\end{equation}
where $\rho_M$ and $\rho_R$ are the matter (including also baryons) and radiation (including neutrinos) energy densities,
$a$ is the scale parameter and $\rho_Q$ is the energy density of the complex quintessence field:
\be{Q(t)}
\rho_Q(t)=|\dot{\psi(t)}|^2+V(\psi (t)).
\ee
The pressure and equation of state of the quintessence fluid are respectively:
\bea{}
p_Q(t) & = & |\dot{\psi(t)}|^2-V(\psi (t))\\
\omega_Q(t) & = &  1- 2{V (\psi (t))\over |\dot{\psi(t)}|^2+V(\psi (t))}.
\label{eqsta}
\eea

In terms of the conformal time $\eta\equiv \ln(a)$
the Friedmann equation can be rewritten as
\be {frie2}
\frac{H^2}{H_0^2}=
\big(\Omega_M e^{-3\eta}+\Omega_R e^{-4\eta}+\frac{\phi'^2}{\rho_C}
+\widetilde{V}(\phi)\big),
\ee
where we have exploited the equation $\dot{\theta}=L/(a^3\phi^2)$ and defined
\be {tildepot}
\widetilde{V}(\phi)=
\frac{L^2}{\phi^2e^{6\eta}\rho_C}+\frac{V(\phi)}{\rho_C}\nonumber.
\ee
As usual, we define the components of the energy density by $\Omega_X=\rho_X/\rho_C$ where $X$ stands for a particular cosmic fluid. 

The equation of motion of the $\phi$ field in terms $\eta$ is
\begin{equation}
0=H^2\phi''+(3H^2+HH')\phi'-\frac{L^2}{\phi^3e^{6\eta}}+\frac{dV(\phi^2)}{d\phi^2}\phi,
\end{equation}
where the derivative of potential reads
\be{derpot}
\frac{\partial V_\phi (\phi^2)}{\partial \phi^2}={\cal{M}}^{\alpha+4}
\Big(\frac{\beta}{2} (\frac \kappa 2 \phi^2)^{\beta/2}-\frac{\alpha}{2}\Big)
\frac{e^{(\frac \kappa 2 \phi^2)^{\beta/2}}}{\phi^{\alpha+2}}.
\ee

The effect of the complex phase $\theta$ in $\psi$ has a different character in the dynamical equations than it has in the equation of state.
In (\ref{eqsta}) it appears as a part of the kinetic term 
in contrast to the equation of motion (\ref{dyna}), where it appears
as a part of the potential contribution.
The extremely steep shape of our potential allows a wide range of the 
field initial values (and  the initial values of the field derivative) to 
develop to a common state, so that a flat and properly accelerating cosmology 
is obtained.

\section{Numerical results and discussion}

We fit our model to the combined Gold + Silver Sn Ia 
dataset from \cite{Riess:2004nr}. This set includes 186 
high red shift supernovae up to $z=1.75$.  

The physical potential parameters, $\cal{M}$ and $L$, are scaled to be dimensionless:
\be{paras}
\mathcal{B}=\frac{L^2}{M_{Pl}^2 \rho_C}, \ \ 
\mathcal{A}=\frac{\mathcal{M}^{\alpha+4}}{M_{Pl}^{\alpha} \rho_C}
\ee
and we use natural units $G_N=M_{Pl}^{-2}=1$. 
The set of model parameters is then
$({\cal{A}(\cal{M})},{\cal{B}}(L),\alpha,\beta,\phi,\phi_{i},\phi'_{i})$. 
For each set of $\beta, \alpha, {\cal{B}}$ and $\phi_i$ the flattest ${\cal{A}}$ is found.
The parameters ${\cal{B}}$ and $\phi_i$ are sampled logarithmically even
and $\alpha$ and $\beta$ with integer steps.
The range of ${\cal{A}}$ corresponds to  physical energy
scales ${\cal{M}}\in (10^{-12},10^{12})$ GeV.
For simplicity, we have fixed the initial value 
of $\phi'_i$ to 0.001. We also looked for an effect of varying $\phi'_i$ within 
the limits $\phi'_i \in (0.00001, 0.1)$, but none was found.
The parameter ranges used in the analysis are shown in the 
table~\ref{tabprms}.

\begin{table} [ht]
\begin{center}
\begin{tabular}{|l|c|}
\hline
 & range \\
\hline
$\beta$ & $0 - 10$\\
$\alpha$ & $0 - 15$\\
$\cal{A}$  & $10^{0} - 10^{-10}$ \\
$\cal{B}$  & $10^{-1} - 10^{-50}$ \\
$\phi_i$ & $0.01 - 1.0$ \\
$\phi'_i$ & $0.001$ \\
\hline
\end{tabular}
\end{center}

\caption{Parameter ranges of the numerical analysis.} \label{tabprms}
\end{table}

The cosmological parameters $h=0.72, \Omega^0_M= 0.27$ and $\Omega^0_R=10^{-5}$
are fixed to be consistent with the current WMAP best fit model data 
\cite{Bennett:2003bz}.
Only flat enough models,
$|1-\Omega_R-\Omega_M-\Omega_Q| \leq 0.02$, are considered.

\begin{figure}[th!]
\leavevmode
\centering 
\vspace*{68mm}
\begin{picture}(0,60)(0,500)
\put(-95,445){$\alpha$}
\put(5,315){$\beta$}
\includegraphics{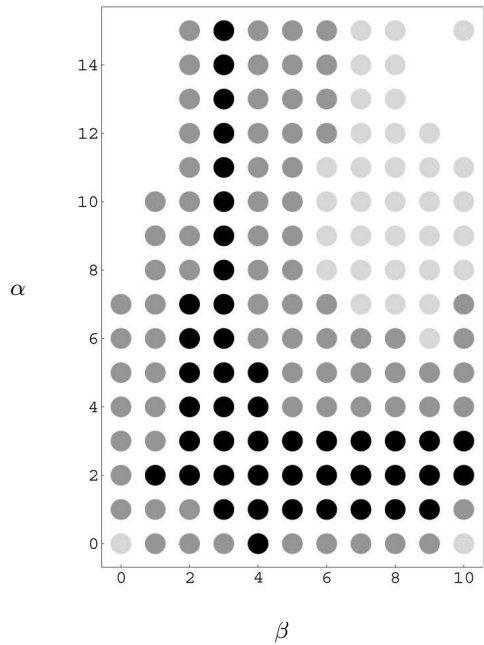}
\end{picture}
\caption{
The fit of the model to the SN Ia dataset when marginalized over 
$\mathcal{B}$ and $\phi_i$ and binned and plotted on the 
($\beta,\alpha$)-plane. 
Only flat solutions within 1, 2 and 3$\sigma$ confidence levels are shown.
1$\sigma$ is depicted with black, 2$\sigma$ with dark gray and 3$\sigma$ 
with light gray circles.
Here 1$\sigma$ covers the boomerang shaped area that continues outside 
the grid, 
although with a constantly decreasing likelihood per bin.}

\label{albe}
\end{figure}

For each point in the parameter space for which $\cal{A}$ can be chosen
so that the universe is flat enough, we fit the SN Ia data to find the
associated $\chi^2$.
The likelihoods are then calculated as usual by assuming gaussian prior 
distributions ($P(X)=\sum_ie^{-\chi^2/2}$).
The data is binned and marginalized to find the confidence levels. 
The results are presented on a ($\beta,\alpha$) parameter plane in 
Fig.~\ref{albe} with normalized likelihoods.
 
From the Fig.~\ref{albe}, it is clear 
that the best likelihoods per bin, depicted by the black points
are clearly concentrated onto a special parameter area.
The 1$\sigma$ area covers a boomerang shaped area that 
continues to very large powers, even outside the considered sensible upper 
boundaries for $\beta$ and $\alpha$ (however, with a constantly decreasing likelihood). 
The 2$\sigma$ area covers almost all the grid with fits 
approximately in between $174\lesssim \langle \chi^2_{bin}\rangle \lesssim 178$.

Some of the represented potentials fit extremely well to the current supernova 
data. Note that the best fitting potentials 
require that the $\beta$ parameter to be 
nonzero. This indicates that the supergravitational ingredient
is important in constructing quintessence dark energy models.

No preferred ${\cal{B}}$ or ${\cal{A}}$ were found.
Although the absolutely best fits of our model had ${\cal{B}}> 10^{-10}$,
these values in general are out of the 2$\sigma$ area. 
The 1$\sigma$ area is restricted to a very small ${\cal{B}}< 10^{-20}$ 
({\it i.e.} $L<10^{-14}$ GeV$^3$).

The $\Lambda$CDM model  or the well known quintessence potentials do not fit 
the data 
particularly well. 
This is easily seen in the Fig.~\ref{albe}.
The $\Lambda$CDM model sits in the origin in the ($\beta,\alpha$) plane,
with $\chi^2_{\Lambda CDM}= 179.3$.
The inverse power model with a complex field lies 
in the $\beta=0$ -axis, and the pure exponential potential in the
$\alpha=0$ -axis.
The BM model (with $\langle \chi^2_{bin}\rangle =174.6$) fits better to the SN Ia data than  
the $\Lambda$CDM and the pure exponential in general
($\beta=4, \alpha=0$ case being 
slightly better).
The  best $(\beta,\alpha)$ bin, is situated at $(3,4)$ with the  
$\langle \chi^2_{bin}\rangle =172.6$.
The absolutely best $\chi^2$ value 171.2 was found at
$\beta=5, \alpha=15, {\cal{B}}=10^{-7}$ ({\it i.e.} $L\sim 10^{-8}$ GeV$^3$), $ \phi_i=0.025,
{\cal{A}}=10^{-9}  $ and $\omega_Q^0=-0.999 $.

\begin{figure}[th!]
\centering 
\psfragscanon
\psfrag{xlabel}{$\alpha$}
\psfrag{ylabel}{$\frac{\cal{M}}{GeV}$}
\includegraphics[width=0.5\textwidth]{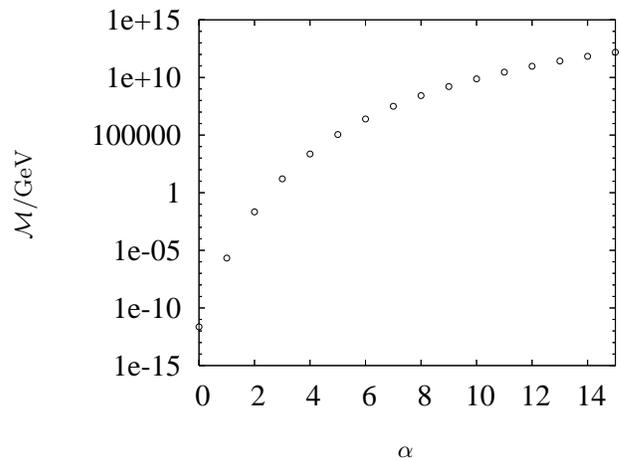}
\psfragscanoff
\caption{The dependence of the energy scale ${\cal{M}}$
on $\alpha$.
The mariginalized data from the fit is fully degenerate in the $\beta$ direction.
Combining this with the Fig.~\ref{albe}, one may conclude that smaller 
energies than $\sim$ TeV are preferred by the fit.
}
\label{enscale}
\end{figure}

The effective mass scale ${\cal{M}(\cal{A})}$  
is fully degenerate in the $\beta$ direction when the 
numerical data is mariginalized over ${\cal{B}}$ and $\phi_i$.
The mass scale is shown in Fig. \ref{enscale} as a function of $\alpha$.
Combining this information with Fig.~\ref{albe}, where
larger values for the inverse power part than $\alpha\geq 4$ restrict $\beta$ to be 3,
 it is evident that the fit somewhat prefers a low effective mass scale,
{\it i.e.} ${\cal{M}(\cal{A})}\lsim 1$ TeV. Put another way, 
given a high effective mass scale, $\beta=3$ is a preferred value. 

Previously a supernova fit and the first CMB Doppler peak consistency have
been done in the range $\beta \in (0,10), \alpha\in (1,10)$  for a  real scalar field
and with an older Sn Ia dataset \cite{Corasaniti:2001mf}
(the most distant supernova in this set was at $z=0.83$). 
Also, the cases $\beta=2$ and $ \alpha=6,11$ have been studied separately in the light 
of CMB data \cite{Brax:2000yb}.
Our results are well consistent with the \cite{Corasaniti:2001mf} Sn Ia 
results. With this more restricted model and generally better dataset, our 
analysis results in a more stringent 1$\sigma$ area for the parameter space.

\section{Conclusions}
In this paper, we have described a supergravitational tracking quintessence 
potential with a complex scalar field, and performed a fit to the 
recent supernova data. 
The studied model can easily fit the data better than the $\Lambda$CDM 
model and the study suggests a combination of inverse power and exponential 
forms for the quintessence potential.

SN Ia data provides 
constraints 
for the studied general
quintessence potential shape.
According to our analysis, for $\alpha \in(0,4)$ the exponent of the potential must 
be $\beta\geq1$ and if $\alpha\geq 4$ then it is required that $\beta = 3$.
The best fit values for the two main model parameters are $\beta= 3$, 
$\alpha=4$.
The effective energy scale $\cal{M}$ proves to be totally degenerate with
respect to the parameter $\beta$. 
We find that a  relatively low energy scale (${\cal{M}}\lsim 1$ TeV) is
favoured by the analysis. Conversely, if a high
energy scale is required, our results srongly prefer $\beta=3$.

The complex contribution is practically negligible 
as effectively all the solutions are found for very small $L$.
However, the best $\chi^2$ value 
($\chi^2=171.2$) is found for an exceptionally high $L$ outside
the 1$\sigma$ area, indicating that in principle the complex part
of the field can play an important role.


Comparing to other commonly considered models, we find that a 
composite potential is strongly preferred by the data.
The fit of the $\Lambda$CDM model is very poor 
(with ${\cal{B}}=0$ and $\chi^2_{\Lambda CDM}=179.3$)
when compared to the other models under study and 
it lies well outside the 3$\sigma$ contour.
The BM model with $\beta=2$, $\alpha=11$ lies within the 2$\sigma$ area.
The equation of state for this case with a small complex contribution
is higher than suggested in \cite{Brax:1999gp}, but
substantially smaller (down to $\omega_{BM}^0=-0.99$) with a sizeable $L$.
The sole inverse power potential does not fit well to the Sn data
and the pure exponential potential proves to model the data comparatively well
only with $\beta=4$. 

Current SN Ia data is accurate enough to distinguish between
tracking quintessence models, given a class of physically motivated
potentials. Within the class of supergravity
inspired potentials, our analysis suggests that
 a composite potential
is preferred over pure exponential or inverse power potentials.


\subsection*{Acknowledgements}
This project has been partly funded by the Academy of Finland project no. 8210338.
KH would like to thank E. Valtaoja for discussions. TM is supported by the 
Academy of Finland.


\end{document}